\documentclass[hyper,12pt]{JHEP3}
\usepackage{epsfig}
\usepackage{amsmath}
\usepackage{amssymb}
\usepackage{epsf}
\usepackage{graphicx}
\usepackage{cite}
\usepackage{graphics}
\usepackage{epstopdf}

\newcommand{\sqrtsnn}{\sqrt{s_{_{{\rm N N}}}}}
\newcommand{\psat}{Q_{{s}}}

\newcommand{\dd}{{\rm d}\,}
\newcommand{\A}{{\rm A}}

\newcommand{\dsigij}{  {\dd {\widehat \sigma}^{\rm dir}_{ij} 
                \over {\dd {\bf p_{_\perp}} \dd y}}}
\newcommand{\dsigijk}{  {\dd {\widehat \sigma}^{\rm frag}_{ij, k} 
                \over {\dd {\bf p_{_\perp}} \dd y}}}
\newcommand{\kd}{  K^{\rm dir} }
\newcommand{\kb}{  K^{\rm frag} }

\newcommand{\alfspi}{ {\alpha_s(\mu) \over 2 \pi} }

\newcommand{\alfas}{ {\alpha_s} }

\newcommand{\be}{\begin{equation}} \newcommand{\ee}{\end{equation}}
\newcommand{\bea}{\begin{eqnarray}} \newcommand{\ena}{\end{eqnarray}}

\def\cO#1{{{\cal{O}}}\left(#1\right)}
   
\def\cO#1{{{\cal{O}}}\left(#1\right)} \def\pt{$p_{_T}$}

\def\kt{{k_{_\perp}}} \def\pt3{{p_{_T{_3}}}}
 
\def\cO#1{{{\cal{O}}}\left(#1\right)} \def\pt{$p_{_T}$}
\def\pt{p_{_\perp}}

\def\z{z_{_{\gamma, \pi}}}
\def\zstar{z^*_{_{\gamma, \pi}}}
\def\zz{z_{_{\gamma, \pi}}}

\title{Quenching of photon and pion spectra \\ at intermediate RHIC energy}

\author{
Fran\c{c}ois Arleo\footnote{On leave from Laboratoire d'Annecy-le-Vieux de Physique Th\'eorique (LAPTH), UMR 5108 du CNRS associ\'ee \`a l'Universit\'e de Savoie, B.P. 110, 74941 Annecy-le-Vieux Cedex, France}\\~\\
CERN, PH department, TH division\\
1211 Geneva 23, Switzerland\\~\\
E-mail: \email{arleo@cern.ch}
}
\abstract{
Single prompt photon and pion spectra in $p$--$p$ and Au--Au collisions at intermediate RHIC energy, $\sqrtsnn=62.4$~GeV, are computed at large $\pt$ in perturbative QCD. Next-to-leading order calculations in $p$--$p$ scattering are first presented. The quenching of the prompt photon and pion yield due to energy loss processes in central Au--Au with respect to $p$--$p$ collisions is then predicted. At this energy, the small phase-space available to produce hard partons makes the pion quenching almost as pronounced as at $\sqrtsnn=200$~GeV, despite the smaller gluon density of the produced medium. In the photon sector, energy loss effects prove small because of this very phase-space restriction, which favours the direct production channel. A significant suppression of high-$\pt$ photons is however predicted, because of a strong isospin effect together with the depletion of nuclear parton densities at large $x$.
}


\preprint{\sffamily{CERN-PH-TH/2007-100}\\\sffamily{LAPTH-1188/07}\\\sffamily{0706.1848v1 [hep-ph]}}

\begin{document}

\setcounter{footnote}{0}
\renewcommand{\thefootnote}{\arabic{footnote}}

\section{Introduction}

Undoubtedly, the strong quenching of leading hadrons up to $\pt\simeq 20$~GeV in central Au--Au collisions at $\sqrtsnn=200$~GeV is one of the most significant results obtained so far at RHIC~\cite{Adcox:2001jpAdler:2002xwAdler:2003qi}. Remarkably, these data appear to be consistent with the ``jet-quenching scenario''~\cite{Bjorken:1982tuGyulassy:1990ye}, where the high-$\pt$ hadron depletion is due to energy loss processes coming from the parton multiple scattering in a dense QCD medium~\cite{Baier:1997krZakharov:1996fvGyulassy:2000fsWiedemann:2000za}. Despite these spectacular experimental results, however, our current theoretical understanding of the fragmentation process modified in QCD media is still not clear. Consequently, there is an important need, on the phenomenological side, to investigate possible energy loss effects on a variety of observables. 

Among others, the production of prompt photons is a particularly interesting probe (see e.g.~\cite{Arleo:2003gn}). For quite some time, the usual belief had been that prompt photons, because of colour neutrality, should not be affected by the dense medium produced in high-energy nuclear collisions and, hence, could serve as a reliable baseline for ``coloured'' hard probes. This, however, may not be correct, and was first called into question a few years ago by Jalilian-Marian, Orginos and Sarcevic~\cite{Jalilian-Marian:2000qa}. Indeed, in QCD perturbation theory at leading-order, prompt photons are produced in the partonic process, as in the Drell-Yan mechanism for instance, but also by the collinear fragmentation of hard quarks and gluons~\cite{Aurenche:1998gv}, just like any leading hadron. There is thus no reason, a priori, why the dense medium responsible for the significant quenching of hadrons should not affect the latter channel as well. As a matter of fact, the PHENIX collaboration recently reported on a preliminary measurement of the prompt photon quenching factor in central Au--Au collisions at $\sqrtsnn=200$~GeV~\cite{Isobe:2007ku}, which may be compatible with energy loss effects in the photon sector~\cite{Arleo:2006xb}. 

Another promising possibility to understand the dynamics of dense medium formation is to investigate, in a systematic way, the energy dependence of the quenching phenomenon~\cite{d'Enterria:2005cs}. In particular, the preliminary data in Au--Au collisions at $\sqrtsnn=62.4$~GeV, together with the ongoing analysis of $p$--$p$ collisions at the same energy (Run-6), may actually provide a missing step between SPS~\cite{d'Enterria:2004ig} and RHIC results~\cite{Adcox:2001jpAdler:2002xwAdler:2003qi}.

The aim of this paper is therefore to investigate the quenching of prompt photons (as well as that of hard pions for consistency) at intermediate RHIC energy, $\sqrtsnn=62.4$~GeV. Predictions at this energy will be given in Section~\ref{se:62} and compared with estimates at top RHIC energy, $\sqrtsnn=200$~GeV. Before discussing these results, we first detail the model used to compute photon and pion spectra in $p$--$p$ and Au--Au collisions in the next Section. 

\section{Model}

\subsection{Perturbative QCD framework in $p$--$p$ collisions}
\label{sec:pp}

Single-inclusive pion and photon hadroproduction cross sections in $p$--$p$ collisions are computed in QCD at next-to-leading order (NLO) in $\alfas$, using the work of Refs.~\cite{Aurenche:1999nz,Aurenche:1998gv,Aurenche:2006vj}:
\begin{eqnarray}
\label{eq:pisingle}
{\dd \sigma^{\pi} \over \dd {\bf p_\perp} \dd y} &=& \sum_{i,j,k=q,g} \int \dd x_1 \ \dd x_2 \
F_{i/p}(x_1, M) \ F_{j/p}(x_2,M) \ {\dd z \over z^2} \ D_{\pi/k}(z,M_F) 
\nonumber \\
&&{}\times \left( {\alpha_s (\mu ) \over 2 \pi} \right)^2
\left [ {\dd \widehat{\sigma}_{ij, k} \over \dd {\bf p_\perp} \dd y}
+ \left( {\alpha_s(\mu) \over 2 \pi} \right) K_{ij,k}(\mu , M, M_F)
\right ],
\end{eqnarray}
for pions, and
\begin{eqnarray}
{\dd \sigma^{\gamma} \over \dd {\bf p_\perp} \dd y}  &=& \sum_{i,j=q,g} \int \dd x_1 \ \dd x_2 
\ F_{i/p}(x_{1},M)\ F_{j/p}(x_{2},M) \nonumber  \\
&\ & \qquad \times
\left( \alfspi \right) \left[ \dsigij \ + \ \left( {\alpha_s(\mu) \over 2 \pi} \right) \ \kd_{ij} (\mu,M,M_{_F}) \right] \nonumber \\
 & + & \sum_{i,j,k=q,g} \int \dd x_{1} \ \dd x_{2}
\ F_{i/p}(x_{1},M)\ F_{j/p}(x_{2},M)\ { \dd z \over z^2} \ D_{\gamma/k}(z, M_{_F})
\nonumber \\
&\ & \qquad \times
\ \left( \alfspi \right)^2\  \left[ \dsigijk \ + 
\ \left( \alfspi \right) \kb_{ij,k} (\mu,M,M_{_F}) \right],
\label{eq:gasingle}
\end{eqnarray}
for photons. In the latter case, we marked the distinction between the photons produced directly (labelled ``dir'') in the partonic scattering process, and those produced from the collinear fragmentation of the hard parton $k$ (labelled ``frag'').

In Eqs.~(\ref{eq:pisingle}) and (\ref{eq:gasingle}), $F_{i/p}$ stands for the parton distribution functions (PDF) of flavour $i$ in a proton, $D_{\pi/k}$ ($D_{\gamma/k}$) for the fragmentation functions (FF) of the parton $k$ into a pion (photon), and $\widehat{\sigma}_{ij, k}$ (respectively, $K_{ij,k}$) for the leading-order (respectively, next-to-leading order) partonic cross section computed in the $\overline{{\rm MS}}$ scheme\footnote{We omit the explicit dependence of $\widehat{\sigma}_{ij, k}$ and $K_{ij,k}$ on the kinematic variables $x_1$, $x_2$, ${\sqrt s}$, $\pt$, and $y$ for clarity.}. We denote by $\mu$, $M$ and $M_{F}$ the renormalisation, the factorisation and the fragmentation scale. 

In this study, we use the NLO CTEQ6M parton density in a proton~\cite{Pumplin:2002vw}, and the NLO Kniehl-Kramer-P\"otter (KKP)~\cite{Kniehl:2000fe}\footnote{Albino, Kniehl, and Kramer (AKK) provide a more recent set of fragmentation functions~\cite{Albino:2005me}. We use, however, the KKP fit for consistency with Ref.~\cite{Arleo:2006xb}.} and the NLO Bourhis-Fontannaz-Guillet (BFG)~\cite{Bourhis:1997yuBourhis:2000gs} fragmentation functions into a pion and into a photon, respectively. Following the scale-fixing procedure used in~\cite{Arleo:2006xb}, scales are chosen so as to minimise the scale-dependence of the NLO predictions. It turns out that rather low scales fulfil this requirement. We choose in the following the range $\pt^2 /8 \le \mu^2 = M^2 = M^2_{_F} \le \pt^2$, both for pions and photons, as a measure of the uncertainty of the NLO calculations.

In the following, we shall be mostly interested in the nuclear effects on single pion and photon $\pt$-spectra in the $5\%$ most central Au--Au collisions. Therefore, only quenching factors
\begin{equation}\label{eq:qf}
R^{\gamma, \pi}(p_{_\perp}) = \frac{1}{N_{_{\rm {coll}}}} \,\, \frac{\sigma_{_{{\rm NN}}}}{\sigma^{{\rm geo}}_{_{{\rm Au Au}}}} \,\times\, {d\sigma_{_{\rm {Au~Au}}}^{\gamma, \pi} \over d{\bf p_\perp} \dd y} \biggr/  {d\sigma_{_{p p}}^{\gamma, \pi} \over d{\bf p_\perp} \dd y},
\end{equation}
are considered, where $\sigma^{{\rm geo}}_{_{{\rm Au Au}}}$ is the geometric cross section obtained via the Glauber multiple scattering theory, $\sigma_{_{{\rm NN}}}$ the nucleon--nucleon cross section, and $\langle N_{{\rm coll}} \rangle \big |_{_{\cal C}}$ the number of binary nucleon--nucleon collisions in central Au--Au collisions~\cite{Arleo:2003gn}. The ratio is properly normalised so that any ``nuclear'' effect (in a broad sense) in Au--Au scattering would make the quenching factor deviating from 1. Moreover, because of the present lack of understanding of medium-modified fragmentation processes, Au--Au spectra (and thus $p$--$p$ spectra in Eq.~(\ref{eq:qf})) are determined at leading-order accuracy. 

\subsection{Nuclear effects in Au--Au collisions}
\label{se:nucleareffects}

Several nuclear effects need to be considered when computing pion and photon large-$\pt$ spectra in Au--Au collisions. The most obvious one concerns the replacement of the proton PDF in Eqs.~(\ref{eq:pisingle}) and (\ref{eq:gasingle}) by:
\begin{equation}\label{eq:iso}
F_{i/\A}(x,M) = Z \, F_{i/p}(x,M) + (A-Z) \, F_{i/n}(x,M),
\end{equation}
for a nucleus with $Z$ protons and $A-Z$ neutrons. The neutron PDF $F_{i/n}$ in Eq.~(\ref{eq:iso}) is obtained from the proton one, $F_{i/p}$, by isospin conjugation: $u^p = d^n$, $d^p = u^n$, $\bar{u}^p = \bar{d}^n$, $\bar{d}^p = \bar{u}^n$, $\bar{s}^p = \bar{s}^n$, and $g^p = g^n$. 

In addition to the above trivial isospin effect, nuclear parton densities differ from those probed in proton targets, over the whole $x$ range~(see~\cite{Armesto:2006ph} for a recent review). 
In order to take into account such a ``genuine'' nuclear effect (as opposed to the isospin corrections), the parton densities in a nucleus are quite generally given by
\begin{equation}\label{eq:shadowing}
F_{i/\A}(x,M) = Z \, F_{i/p}(x,M) \, S^\A_{i/p}(x,M) + (A-Z) \, F_{i/n}(x,M) \, S^\A_{i/n}(x,M),
\end{equation}
in the perturbative calculations.  Here, $S^A_{i/p}$ ($S^A_{i/n}$) denotes the ratio of the proton (neutron) PDF in a nucleus $\A$ over that in a ``free'' proton (neutron). It is given throughout this paper by a global fit of Drell-Yan and DIS data on nuclear targets (EKS,~\cite{Eskola:1998df}). Note that at very small $x$, however, coherence effects become large and the above factorised form for the nuclear PDFs may no longer be valid. Nevertheless, the typical $x$ values we shall meet in this study are pretty large, say $x \gtrsim 0.01$, hence the use of Eq.~(\ref{eq:shadowing}) appears to be fully justified.

Finally, the partonic energy loss process in the dense medium is addressed. We shall follow a standard procedure, and model the energy loss through a shift in the $z$ variable entering the fragmentation functions,~\cite{Wang:1996yh}
\begin{equation}
  \label{eq:modelFF}
\zz\,D_{\gamma,\pi/k}^{\rm med}(\zz, M_{_F}, \kt) = \int_0^{\kt (1
  - \z)} \, \dd \epsilon \,\,{\cal P}_k(\epsilon, \kt)\,\,\,
\zstar\,D_{\gamma,\pi/k}(\zstar, M_{_F}),
\end{equation}
${\cal P}_d(\epsilon, \kt)$ standing for the probability that the hard parton $k$ with transverse momentum  $\kt$ has lost an energy $\epsilon$. It was introduced in~\cite{Baier:2001yt} and computed numerically in~\cite{Arleo:2002kh} from the Baier--Dokshitzer--Mueller--Peign\'e--Schiff medium-induced gluon spectrum~\cite{Baier:1997sk}. This spectrum --~and thus eventually the quenching factors~-- depends on one scale, $\omega_c$. Accounting for the expansion of the produced medium, it reads~\cite{Salgado:2002cd,Baier:2002tc}
\begin{equation}
  \label{eq:omc}
  \omega_c \simeq \hat{q}(t_0) \ t_0 \ L,
\end{equation}
assuming a purely longitudinal expansion. The initial-time transport coefficient $\hat{q}(t_0)$ in Eq.~(\ref{eq:omc}) represents the scattering power of the medium, and $L$ the length covered by the hard parton in the medium. The transport coefficient has been determined perturbatively in a cold nucleus, $\hat{q}\simeq 0.045$~GeV$^2$/fm~\cite{Baier:1997sk}, in good agreement with a phenomenological estimate from Drell-Yan production off nuclei~\cite{Arleo:2002ph}. In hot quark-gluon plasma, it is expected to scale like~\cite{Baier:2002tc}
\begin{equation}
  \label{eq:tc}
  \hat{q} \sim \alfas \ \epsilon^{3/4},
\end{equation}
where $\epsilon$ is the medium energy-density and $\alfas$ the strong coupling constant evaluated at a scale $\cO{\hat{q} L}$. The energy-density reached in central heavy-ion collisions, estimated in the saturation model of Ref.~\cite{Eskola:1999fc}, shows a parametric dependence $\epsilon \sim \psat^4$, at an initial time $t_0 \sim 1/\psat$, with $\psat$ the gluon saturation-momentum. From Eqs.~(\ref{eq:omc}) and (\ref{eq:tc}), $\omega_c$ should scale like $\psat^2$. From the $\sqrtsnn$ dependence of the saturation momentum~\cite{Eskola:1999fc}, we thus have\footnote{Also interesting is the atomic mass dependence of the $\omega_c$ scale. From~\cite{Eskola:1999fc}, we get $\omega_c \sim A^{0.25} L \sim A^{0.58}$. This could be useful to predict the pion/photon quenching in lighter systems such as Cu--Cu at RHIC or Ar--Ar at the LHC.}
\begin{equation}
  \label{eq:omcvss}
\omega_c (\sqrtsnn) \propto \left(\sqrtsnn \right)^{0.4},
\end{equation}
neglecting the running of $\alfas$ in Eq.~(\ref{eq:tc}). We shall take the $\omega_c = 10$--$15$~GeV range to predict the pion and photon quenching at $\sqrtsnn = 62.4$~GeV, to be consistent with our previous estimate, $\omega_c = 20$--$25$~GeV, at RHIC full energy~\cite{Arleo:2006xb}.

To summarise, quenching factors are computed, in the following, under various assumptions concerning the nuclear effects:
\begin{itemize}
\item[(i)] Only isospin corrections, Eq.~(\ref{eq:iso}), are considered, that is, neglecting any nuclear corrections to the PDFs nor energy loss effects. This prediction should be rather model-independent, and is labelled ``Au Au'' in the figures;
\item[(ii)] Isospin and nuclear PDF (EKS) corrections are included in the nuclear parton densities, Eq.~(\ref{eq:shadowing}).  These predictions are labelled ``Au Au EKS'';
\item[(iii)] Parton densities include isospin/nPDF corrections, Eq.~(\ref{eq:shadowing}), and medium-modified fragmentation functions, Eq.~(\ref{eq:modelFF}), are used to account for energy loss processes (labelled ``Au Au EKS $\omega_c = 10$--$15$~GeV'').
\end{itemize}

Although the above-mentioned mechanisms are expected to be dominant, it is worth to mention that other nuclear effects could affect the quenching of high-$\pt$ particles. It is the case for instance of the initial-state multiple scattering of partons in the nuclei, leading to the well-known Cronin effect~(see e.g.~\cite{Accardi:2002ik} for a review). It is is expected to be significant at not too large $\pt$ (say, below 5~GeV~\cite{Vitev:2004gn}) and is neglected in this study.

\section{Results}
\label{se:62}

\subsection{Absolute spectra}
\label{se:spectra}

\begin{figure}[ht]
  \begin{minipage}[ht]{7.2cm}
    \begin{center}
      \includegraphics[height=7.2cm]{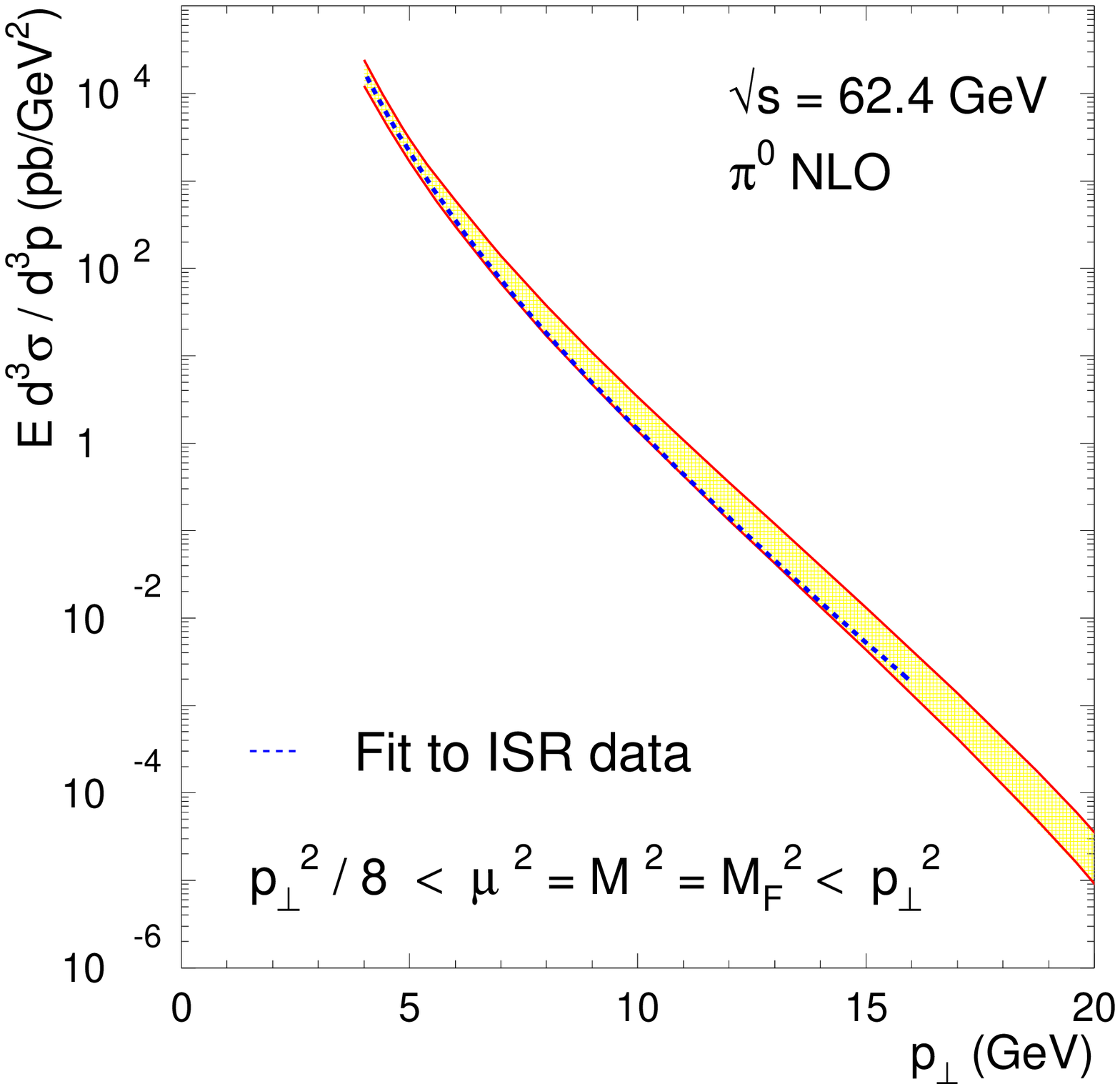}
    \end{center}
  \end{minipage}
~
  \begin{minipage}[ht]{7.2cm}
    \begin{center}
      \includegraphics[height=7.2cm]{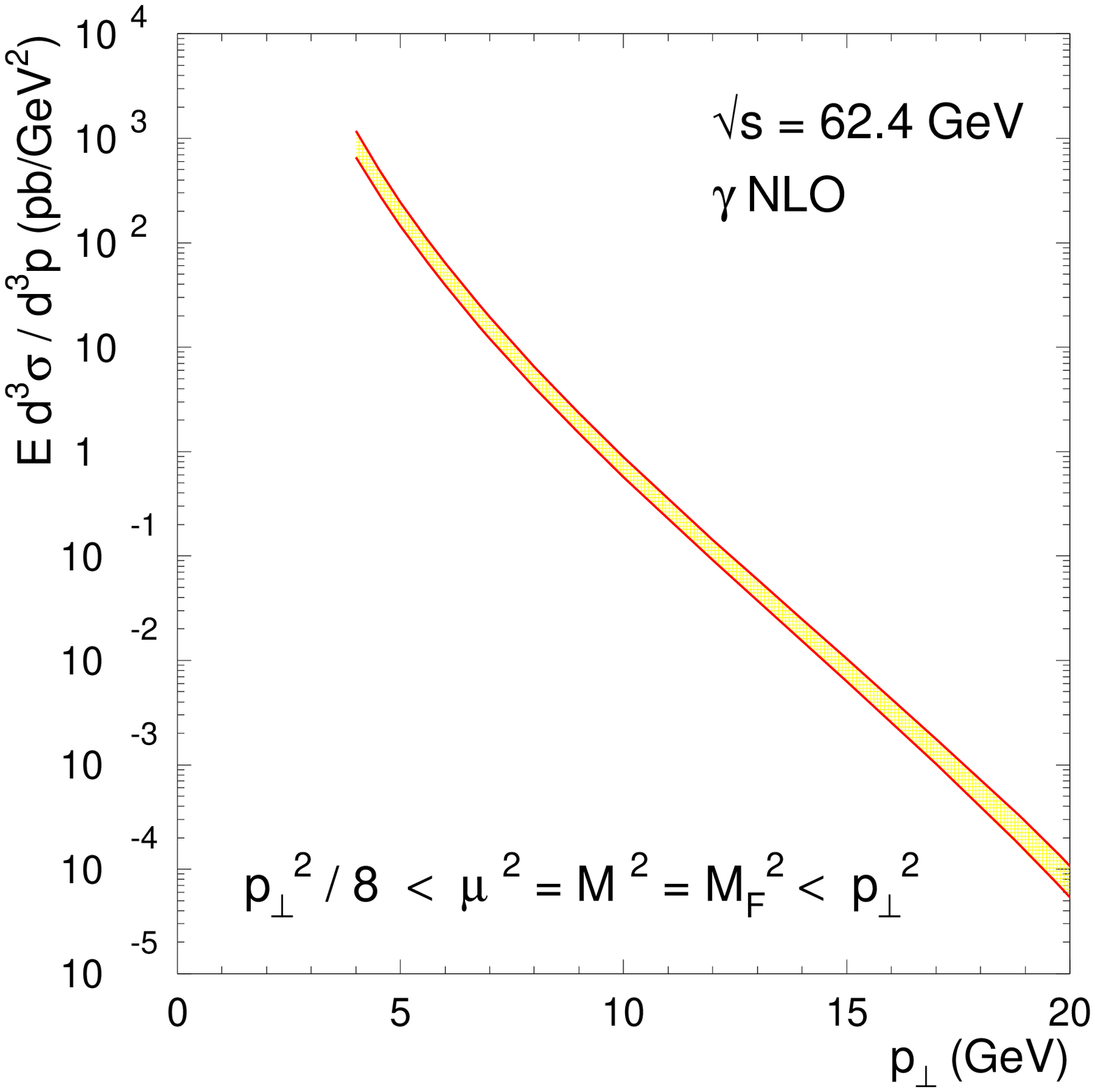}
    \end{center}
  \end{minipage}
  \caption{Pion (left) and photon (right) invariant cross section in $p$--$p$ collisions at $\sqrt{s}=62.4$~GeV computed at NLO accuracy, varying simultaneously the factorisation, the renormalisation and the fragmentation scales from $\pt/\sqrt{8}$ to $\pt$. Pions and photons are produced in the $[-0.35;0.35]$ rapidity interval.}
  \label{fig:spectra}
\end{figure}

The transverse momentum spectra in $p$--$p$ collisions at $\sqrtsnn=62.4$~GeV are shown in Fig.~\ref{fig:spectra} for pions (left) and photons (right). The fragmentation process in the pion channel makes the cross sections decreasing much faster with $\pt$ than what is predicted for the photons, which are produced {\it directly} in the partonic process at large $x_{_\perp} = 2 \pt / \sqrtsnn \gtrsim 0.1$. Both processes turn out to have similar cross sections around $\pt \simeq 15$~GeV.

The theoretical uncertainty in the pion sector is pretty large, from a factor 2 at low $\pt$ up to a factor 3 at $\pt \simeq 20$~GeV. At large $\pt$, most of this uncertainty can be attributed to the fragmentation process (hence to the fragmentation-scale dependence) since fragmentation functions are probed at large $z$. On the contrary, it is interesting to remark that large-$\pt$ photons, being produced directly, show a somewhat lesser scale-dependence (factor of 2 uncertainty over the whole $\pt$ range). 

Pions and photons have been measured a long time ago in $p$--$p$ collisions at $\sqrt{s}=62$--$63$~GeV at the ISR (see~\cite{d'Enterria:2004rp} for a list of references), and subsequent perturbative calculations were shown to describe fairly these data. We refer in particular the reader to Refs.~\cite{Aurenche:1999nz,Aurenche:1998gv,Aurenche:2006vj} for a complete review of single-inclusive pion and photon QCD predictions from fixed-targets to collider experiments. Recently, all ISR $\pi^0$-data were fitted with an empirical law~\cite{d'Enterria:2004rp} reproduced as a dashed line in Fig.~\ref{fig:spectra}. The agreement between this fit and the present calculation is particularly good, although it is worth to remark that the ``data'' are best reproduced\footnote{It should however be kept in mind that the ISR measurements have a rough 20\%-spread around this best fit.} with the present lower estimate, i.e. using all scales equal to $\pt$. 

\subsection{Quenching factors}
\label{se:quenching}

\begin{figure}[ht]
  \begin{minipage}[ht]{7.2cm}
    \begin{center}
      \includegraphics[height=7.2cm]{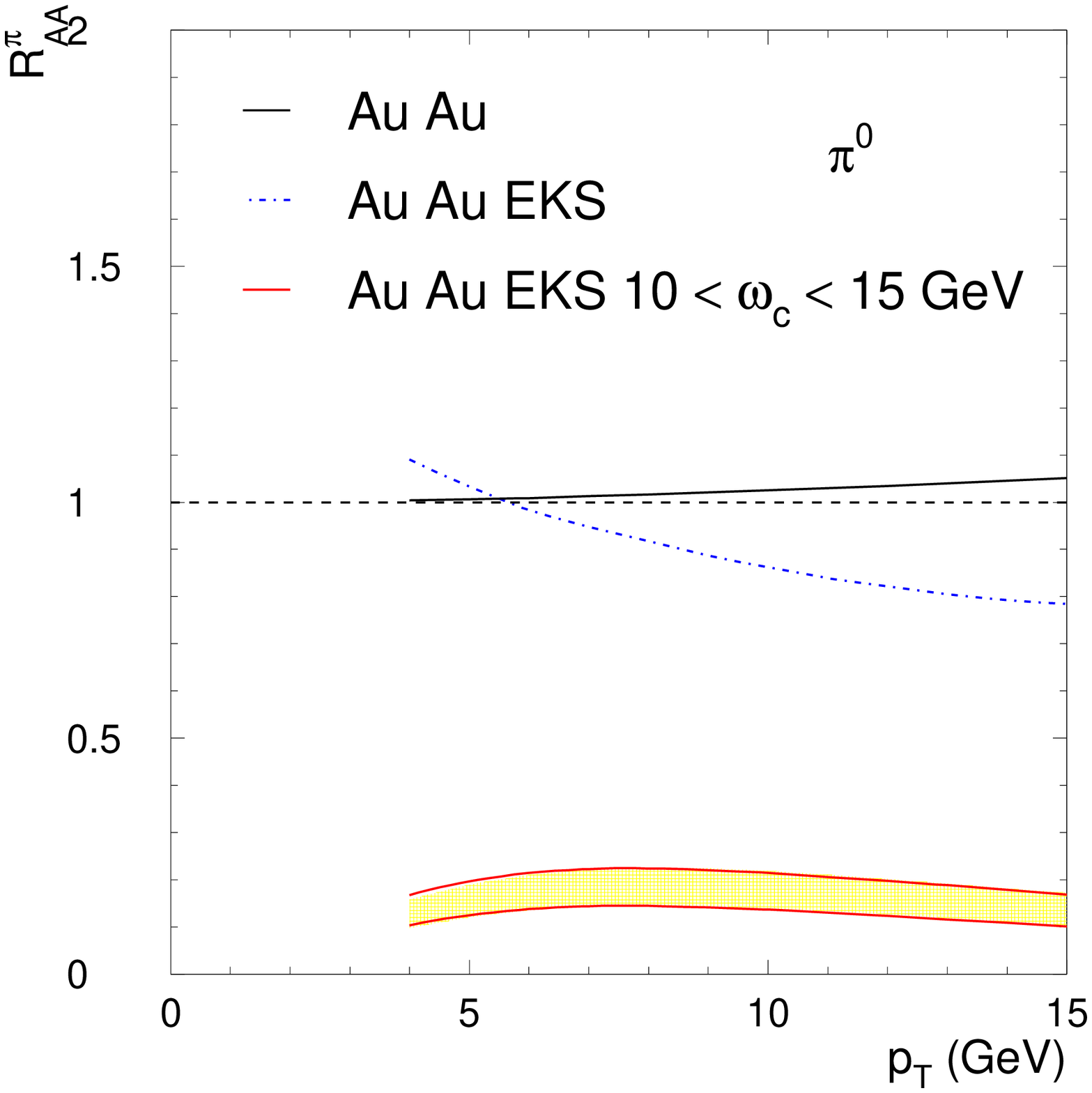}  
    \end{center}
  \end{minipage}
~
  \begin{minipage}[ht]{7.2cm}
    \begin{center}
      \includegraphics[height=7.2cm]{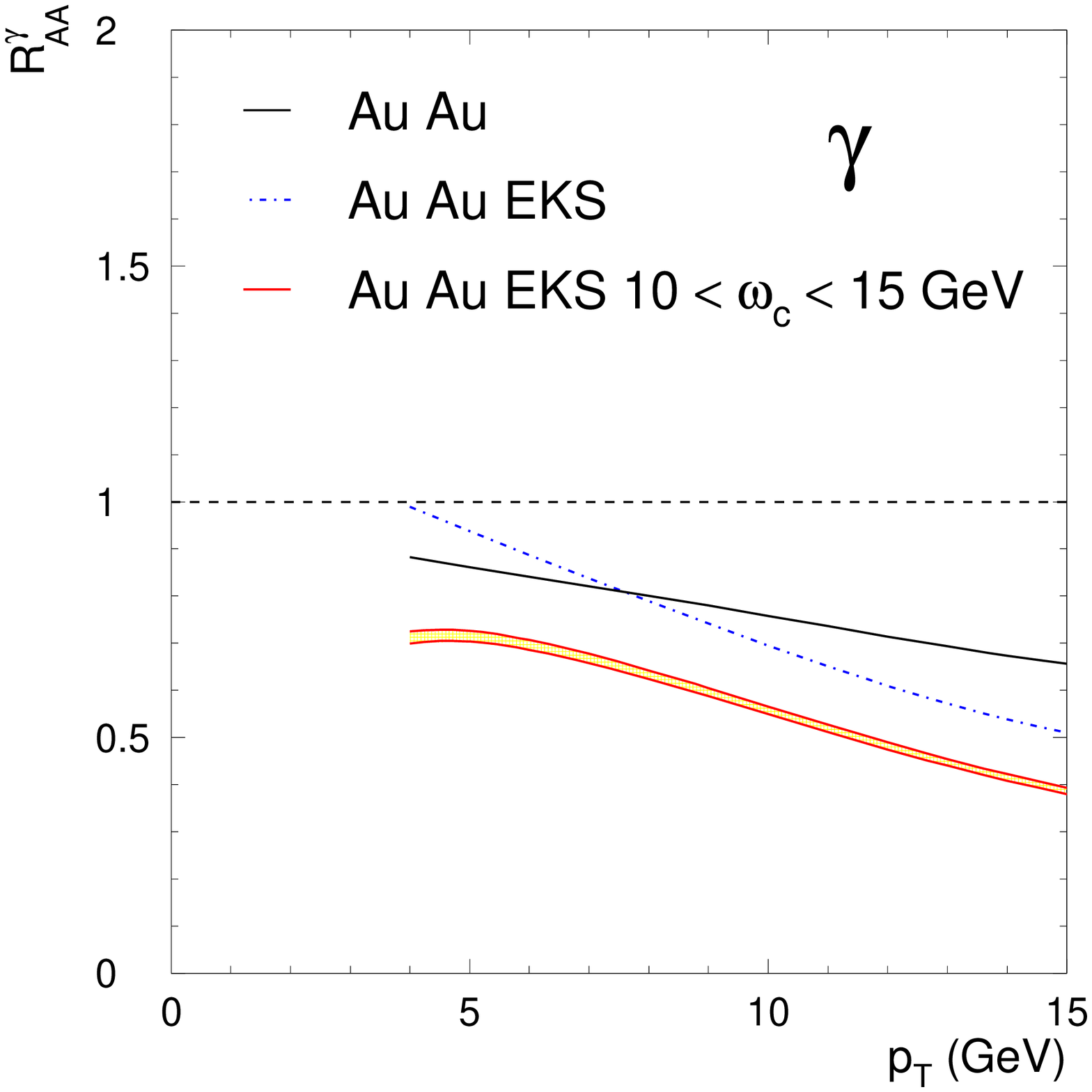}  
    \end{center}
  \end{minipage}
  \caption{Pion (left) and photon (right) quenching factors in central Au--Au collisions at $\sqrtsnn=62.4$~GeV. Calculations are done at LO, assuming (i) isospin (solid), (ii) isospin and nPDFs (dotted), (iii) isospin, nPDFs and energy loss (band) effects (see text for details).}
  \label{fig:quenchingsingle}
\end{figure}

Quenching factors in $0$--$5\%$ central Au--Au collisions at $\sqrtsnn=62.4$~GeV are computed in Fig.~\ref{fig:quenchingsingle} for pions and photons. Let us first comment on the pion case (Fig.~\ref{fig:quenchingsingle}, left). Isospin corrections prove tiny, with a 5\% enhancement ($R_{_{\rm Au Au}} >1$), at most, at high $\pt \simeq 15$~GeV. More pronounced are the nuclear effects in the nPDFs which deplete somehow (20\%) the pion yield in nuclei due to the EMC effect at large $x \gtrsim 0.2$. The slight enhancement seen at $\pt \simeq 5$~GeV is due to the rather strong gluon anti-shadowing assumed in the EKS parametrisation around $x \sim 0.1$. Overall, both isospin and nPDFs corrections remain small and do not affect the pion quenching factor by more than 20\% over the whole $\pt$ range. These prove, however, more pronounced at $\sqrtsnn=62.4$~GeV than at $200$~GeV due to the larger $x$ probed in the nuclear parton densities.

On the contrary, the energy loss mechanism quenches the pion $\pt$ spectrum by roughly a factor of 5! Furthermore, the quenching factor is seen to be rather $\pt$-independent; this flat behaviour coming from the interplay of nPDFs corrections at large $x$ (EMC effect) and energy loss processes. Quite surprisingly, the quenching turns out to be as pronounced as at $\sqrtsnn=200$~GeV, despite the twice smaller $\omega_c$-scale. The reason is actually due to kinematics. With decreasing $\sqrtsnn$ and at a given $\pt$ (i.e. with increasing $x_{_\perp}$), the phase-space available to produce high-$\kt$ partons becomes more and more restricted. The fragmentation process into the pion hence occurs at larger $z$, where FFs are known to fall dramatically, and therefore where energy loss effects should be the strongest. This --~as well as the more pronounced nPDF corrections~-- explains why the pion quenching factor proves as small as $\sim 0.2$ despite the much smaller $\omega_c = 10$--$15$~GeV assumed in the calculation. 

The pion quenching at $\sqrtsnn=62.4$~GeV has been already computed within various parton energy-loss models by Vitev~\cite{Vitev:2004gn}, Adil and Gyulassy~\cite{Adil:2004cn}, Wang~\cite{Wang:2004yv}, and Eskola, Honkanen, Salgado, Wiedemann~\cite{Eskola:2004cr}. All calculations predict slightly larger quenching factors, $R^{\pi}\simeq 0.3$--$0.4$, in better agreement with PHENIX preliminary results in Au--Au collisions~\cite{Buesching:2006ap} normalised to ISR $p$--$p$ data~\cite{d'Enterria:2004rp}. It should be interesting to see whether the final PHENIX $p$--$p$ measurements confirm this first observation. The overestimate of the quenching in the present model at low $\pt \simeq 4$--$6$~GeV has already been pointed out at $\sqrtsnn=200$~GeV in Ref.~\cite{Arleo:2006xb} and is possibly due to the Cronin effect not considered here\footnote{We may also remark that the energy dependence assumed here, Eq.~(\ref{eq:omcvss}), is slower than the $\sqrtsnn^{0.57}$ assumed in Ref.~\cite{Eskola:2004cr}.}. Hopefully future data will be able to tell whether this overestimate subsists at larger $\pt$.

\begin{figure}[ht]
  \centering
  \includegraphics[height=9.cm]{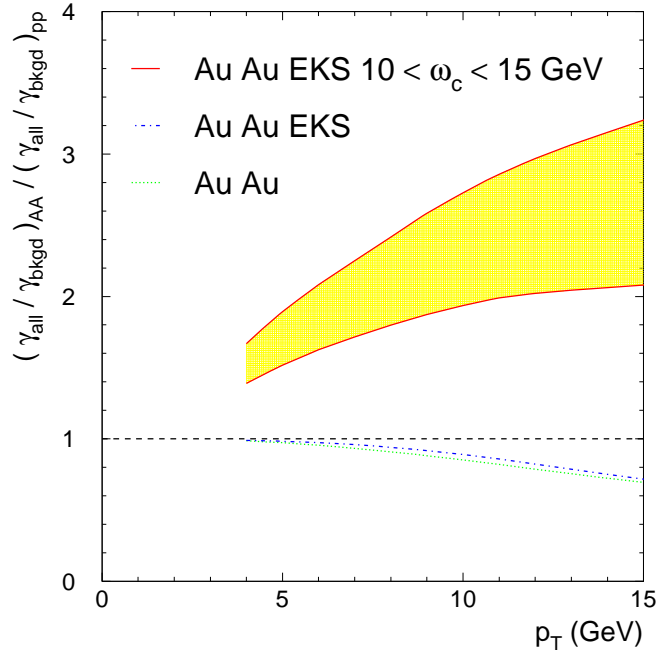}
  \caption{Ratio of the overall photon production over the background, $\gamma_{_{\rm all}}/\gamma_{_{\rm bkgd}}$, in central Au--Au collisions normalised to the $p$--$p$ case. LO calculations are done in $p$--$p$ and Au--Au collisions at $\sqrtsnn=62.4$~GeV.}
  \label{fig:gammaoverpion}
\end{figure}

The quenching pattern in the photon channel (Fig.~\ref{fig:quenchingsingle}, right) is quite different. Already noticeable in the predictions at $\sqrtsnn=200$~GeV~\cite{Arleo:2006xb}, and possibly seen in the PHENIX preliminary measurements~\cite{Isobe:2007ku}, the isospin corrections at $\sqrtsnn=62.4$~GeV now prove spectacular: from $\pt=4$~GeV to $\pt=15$~GeV, the quenching factor decreases from $0.85$ down to $0.65$. Because of the electromagnetic coupling, the scattering of up-quarks is indeed favoured as compared the down-quarks due to their respective electric charge. The relative lack of valence up-quarks in the neutron (as compared to a proton) tends to deplete photon cross section in large nuclei (with $Z \sim A/2$), which explains why the quenching factor is expected to be smaller than 1. Nuclear PDF corrections further increase the negative slope of the photon quenching, making the ratio $R\simeq 1$ at low $\pt$ because of anti-shadowing, and $R\simeq 0.5$ in the highest $\pt$ bin (EMC effect). As emphasised in~\cite{Arleo:2006xb}, energy loss could also affect prompt photon production in Au--Au collisions because of the fragmentation contribution (see Eq.~(\ref{eq:gasingle})). However, its effects remain rather small, roughly 20\% for all $\pt$. As a matter of fact, at low $\sqrtsnn$, photons are mostly produced directly in the hard subprocess and should not be too much quenched by the medium. In that sense, the restricted phase-space has opposite consequences --~as far as energy loss processes are concerned --~in the pion and the photon channel, because of their different dynamical production process. In particular, neglecting any isospin/nPDF corrections, the quenching factor at the very edge of phase-space should be, strictly speaking, $R^\pi(2\pt=\sqrtsnn)=0$ for pions but $R^\pi(2\pt=\sqrtsnn)=1$ for photons.

This difference is well illustrated in the $\gamma_{_{\rm all}}/\gamma_{_{\rm bkgd}}$ ratio, that is the overall photon yield (including the prompt photon signal and the $\pi^0 \to \gamma\gamma$ decay background\footnote{As in Ref.~\cite{Arleo:2006xb}, the $\pi^0$ decay spectrum is estimated to be $2/(n-1)$ that of the single-inclusive $\pi^0$ spectrum, where $n$ represents the power-law exponent of the $\pi^0$ spectrum in $p$--$p$ collisions~\cite{Arleo:2003gn}.}) normalised to the background. This ratio is, however, quite sensitive to the absolute magnitude of the $\pi^0$ and $\gamma$ spectra in $p$--$p$ collisions. In order to lower this uncertainty, Fig.~\ref{fig:gammaoverpion} displays the double ratio $\left(\gamma_{_{\rm all}}/\gamma_{_{\rm bkgd}}\right)_{{\rm Au Au}} \bigr/ \left(\gamma_{_{\rm all}}/\gamma_{_{\rm bkgd}}\right)_{p p}$ as a function of the transverse momentum. This observable looks promising as it is clearly sensitive to energy loss processes --~see e.g. the strong enhancement in Fig.~\ref{fig:gammaoverpion} and its $\omega_c$ dependence~-- while rather independent of isospin/shadowing corrections when going from $p$--$p$ to Au--Au collisions.

\subsection{Comparison with $\sqrtsnn=200$~GeV}

Naively, we could think that the smaller the incident energy $\sqrtsnn$ (and therefore the smaller the medium energy-density), the weaker the quenching. Looking at Eq.~(\ref{eq:omcvss}), this would mean that the quenching factor smoothly decreases with $\sqrtsnn$. This is not necessarily true, for two reasons. The first one comes from the nPDF corrections: parton density ratios, $S^\A_i$, clearly have a non-monotonic behaviour as a function of Bjorken-$x$ or, alternatively, as a function of $\sqrtsnn$ at a given $\pt$. The second reason deals with the phase-space restriction at small $\sqrtsnn \gtrsim \omega_c$, already mentioned in Sect.~\ref{se:quenching}: energy loss effects can be magnified at low $\sqrtsnn$ because of the larger $z$ values probed in the fragmentation process (or equivalently because of the {\it steeper} partonic spectra at large $\kt\lesssim\sqrtsnn/2$).

\begin{figure}[ht]
  \begin{minipage}[ht]{7.2cm}
    \begin{center}
      \includegraphics[height=7.2cm]{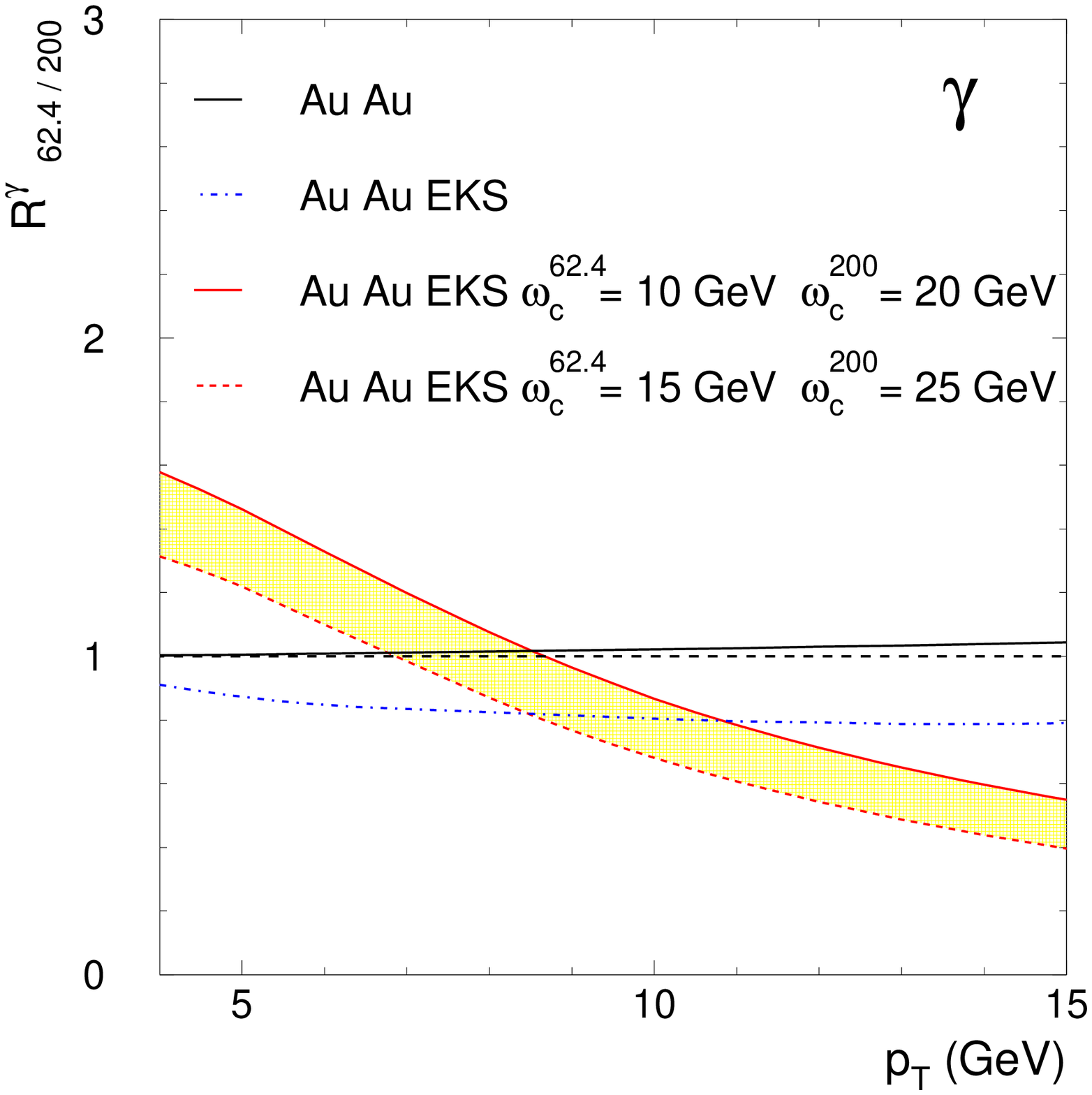}  
    \end{center}
  \end{minipage}
~
  \begin{minipage}[ht]{7.2cm}
    \begin{center}
      \includegraphics[height=7.2cm]{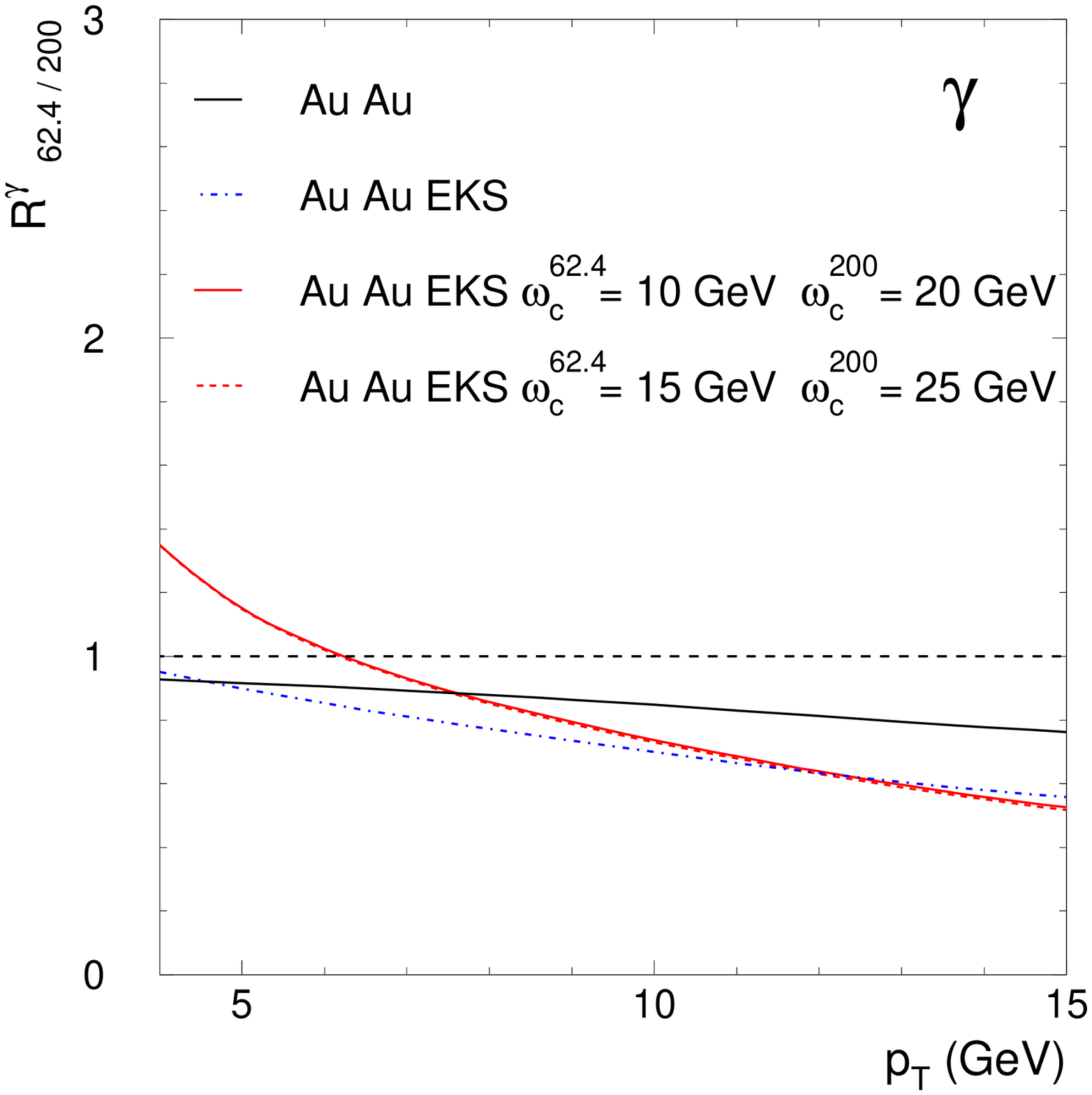}  
    \end{center}
  \end{minipage}
  \caption{Ratio of the quenching factor at $\sqrtsnn=62.4$~GeV over that at $\sqrtsnn=200$~GeV in the pion (left) and the photon (right) channel. Calculations are done at LO, assuming (i) isospin (solid), (ii) isospin and nPDFs (dash-dotted), (iii) isospin, nPDFs and energy loss (band) effects (see text for details).}
  \label{fig:quenching62vs200}
\end{figure}

In order to compare the quenching at intermediate and full RHIC energy, the double ratio
\begin{equation}
  \label{eq:rs}
  R_{~62.4/200}  = \frac{R_{{\rm A A}}(\sqrtsnn=62.4\ {\rm GeV})}{R_{{\rm A A}}(\sqrtsnn=200\ {\rm GeV})},
\end{equation}
is computed in Fig.~\ref{fig:quenching62vs200}. This ratio has been introduced by Adil and Gyulassy in~\cite{Adil:2004cn} in order to get rid of the uncertainty in the initial gluon rapidity-distribution, and was determined in the pion channel assuming energy loss effects only. 

Although the energy loss probability distribution used here is somehow different to the one assumed in Ref.~\cite{Adil:2004cn}, it is worth to mention that roughly the same trend is observed in our model (Fig.~\ref{fig:quenching62vs200}, left), and especially the negative $\pt$-slope of this ratio coming from the phase-space restriction above discussed. Our prediction is in particular similar to their calculation assuming an initial gluon distribution ${\rm d}N_g / {\rm d}y=770$~\cite{Adil:2004cn}. The inclusion of nPDF corrections only affects the magnitude of the ratio, lowering it by 20\% (dashed-line). Consequently, the transverse momentum above which the quenching is larger at $\sqrtsnn=62.4$~GeV than at $\sqrtsnn=200$~GeV ($R_{~62.4/200}(\pt)\le 1$) is $\pt \simeq 8$~GeV instead of $\pt \gtrsim 12$~GeV in Ref.~\cite{Adil:2004cn}.
 
The same calculation is carried out in Fig.~\ref{fig:quenching62vs200} (right) in the photon channel. Despite the different dynamical production process, the ratio $R_{~62.4/200}$ including energy loss is remarkably similar to what is found in the pion sector. This is, however, largely accidental, since the isospin corrections in the double ratio turn out to be really different for pions and for photons. Interestingly, the double ratio is sensitive to energy loss effects at low $\pt\lesssim 7$~GeV. At larger $\pt \gtrsim 11$~GeV, the ratio does not depend much on the energy loss mechanism any longer (because of the direct process) but rather should probe the EMC effect in the nPDFs.

\section{Summary}

Large-$\pt$ pion and photon spectra have been considered in $p$--$p$ and Au--Au scattering at intermediate RHIC energy, $\sqrtsnn=62.4$~GeV. After discussing NLO predictions in $p$--$p$ collisions, quenching factors in the $0$--$5\%$ most central Au--Au collisions are predicted for both pions and photons within the same model. The energy loss process quenches rather strongly the pion yield, but proves rather inefficient for the photons, whose quenching factor is dominated by significant isospin and nPDF corrections. The quenching of the photon total yield over its background is also determined and turns out to be rather sensitive to the energy loss processes.

Predictions at $\sqrtsnn=62.4$ and at $\sqrtsnn=200$~GeV are then compared. Despite the smaller gluon density, the large-$\pt$ pion quenching at $\sqrtsnn=62.4$~GeV proves actually as pronounced as at full RHIC energy because of the strong phase-space restriction at this energy, as already mentioned in~\cite{Adil:2004cn,Eskola:2004cr}. On the contrary, the kinematic dependence of the photon quenching is opposite because of the direct contribution which dominates the inclusive yield: when the phase-space becomes too restricted --~i.e. at low $\sqrtsnn$~--, photons are produced directly and should therefore not be affected by the dense QCD medium.

\section*{Acknowledgements}

I thank warmly Patrick Aurenche and David d'Enterria for many discussions and useful comments as well as for the reading of the manuscript.

\providecommand{\href}[2]{#2}\begingroup\raggedright\endgroup

\end{document}